\documentclass{PoS}

\title{Twisted Poisson Structures and Non-commutative/non-associative Closed String Geometry}

\rightline{\small LMU-ASC 32/12}
\rightline{\small MPP-2012-83}

\ShortTitle{Non-commutative closed strings}

\author{\speaker{Dieter L\"ust}
\\
 Arnold Sommerfeld Center for Theoretical Physics\\
Department f\"ur Physik, Ludwig-Maximilians-Universit\"at M\"unchen\\
Theresienstr.~37, 80333 M\"unchen, Germany       \\
and\\
Max-Planck-Institut f\"ur Physik\\
F\"ohringer Ring 6, 80805 M\"unchen, Germany\\
        E-mail: \email{dieter.luest@lmu.de, luest@mppmu.mpg.de}}

\abstract{In this paper we discuss non-commutative and non-associative geometries that emerge
in the context of non-geometric closed string backgrounds. T-duality and doubled field theory plays an important role in formulating the corresponding effective action
for these kind of non-geometric string backgrounds. As we will argue, the emerging non-commutative and non-associative algebras for the closed string (dual) coordinates and
(dual) momenta can be mathematically described by a twisted Poisson structure, in closed analogy to the phase space of a point particle
moving in the field of a magnetic monopole. }

\FullConference{Proceedings of the Corfu Summer Institute 2011 School and Workshops on Elementary Particle Physics and Gravity\\
		September 4-18, 2011\\
		Corfu, Greece}

\def\R {\mathcal{R}}

\def\tg{\tilde{g}}

\def\b {\beta}
\def\tx{\tilde{x}}

\newcommand{\be}{\begin{equation}}
\newcommand{\ee}{\end{equation}}

\newcommand{\tri}{\hspace{-3.5pt}\vartriangle\hspace{-3.5pt}}

\begin{document}

String flux compactifications (for reviews see e.g. see \cite{Grana:2005jc,Blumenhagen:2006ci})
attracted a lot of interest during the recent years.
In particular, they proved to be essential for moduli stabilization,
they led to the discovery of a huge string flux landscape,
and they are also relevant in the context of string inflationary scenarios
in the early universe.
Moreover, flux compactifications broadened the notion of geometry in string theory,
since it became clear that generalized geometries, that are not
geometric spaces, can still serve as consistent string backgrounds.
Finally, T-duality also plays an important role within flux compactifications:
background with and without $H$-fluxes can be T-dual to each other, and T-duality
can transform a geometric space into a non-geometric background and vice versa.
The so-called doubled geometry, where one includes
the momentum coordinates together with the dual winding coordinates, turned out to be
helpful to formulate an effective string action, which is invariant under T-duality transformations.

In any case, an extended string generically feels the background geometry rather differently compared to
a point particle.     As it was discussed  in the past by several authors, non-commutative geometry 
may naturally arise in string theory. In particular for open strings on the (2-dimensional) world volume of D-branes,
their end-point coordinates become non-commuative when a constant $B$- or $F$-field on the D-brane
is turned on (see \cite{Chu:1998qz,Schomerus:1999ug,Seiberg:1999vs,Ardalan:1998ks} 
for a small selection of papers). In this background, the corresponding open string conformal field theory is still a free
theory and hence exactly solvable.  As a result of the CFT computation, 
the non-vanishing $B$- or $F$-field induces a non-vanishing commutator of the open string end point coordinates of the form (${\cal F}=B-F$)
\begin{equation}\label{commopen}  [ X_1(\tau), X_2(\tau) ]=\theta_{12}\quad{\rm with}\quad \theta_{12}=-{2\pi i \alpha' {\cal F}\over 
      1+{\cal F}^2}
      \end{equation}
        Performing a T-duality transformation along one of the D-brane world volume coordinates,  
  one obtains a geometric configuration with (1-dimensional) D-branes that intersect the coordinates axis in a certain angle, which is determined by the $B$- or $F$-field.
Now, after the T-duality transformation, the dual open string coordinates are fully commuting.

This non-commutative open string algebra corresponds to a {\sl Poisson structure}, which is similar to the Poisson structure of a point particle moving in a constant magnetic field,
where however the non-commutativity shows up in the algebra of the momentum operators.
By evaluating correlation functions of vertex operators
in open string theory, it is possible to 
derive the Moyal-Weyl product.  The algebra of the periodic functions can then be defined by  an $N$-product $\star_N$:
\begin{eqnarray}
  \label{Nbracketcon}
  & f_1(x)\, \star_N\,  f_2(x)\, \star_N \ldots \star_N\,  f_N(x) := \\
  &\hspace{40pt}\exp\left( i  \sum_{1\leq n< m\leq N}
     \theta^{ab}\,
      \partial^{x_n}_{a}\,\partial^{x_m}_{b}  \right)\, f_1(x_1)\, f_2(x_2)\ldots
   f_N(x_N)\Bigr|_{x_1=\ldots = x_N=x} \;.
\end{eqnarray}
These $N$-products are  related to 
the subsequent application of the usual star-product $\star=\star_2$ in the following way
\begin{equation}
   f_1\star_N f_2\star_N \ldots \star_N  f_N=
   f_1\star f_2\star \ldots \star  f_N\; .
\end{equation}

Here we are investigating the  geometry of closed strings moving in an $H$-field background and its T-dual versions.
This is a more difficult problem, since a non-vanishing $H$-field implies
that the $B$-field is non-constant and hence the corresponding two-dimensional $\sigma$-model
is in general not anymore exactly solvable. 
The deformation from the flat space by the $H$-field flux  or, respectively, by its T-dual metric components imply
that we necessarily have to consider at least three-dimensional backgrounds in order to obtain closed 
string non-commutativity, whereas open string non-commutative geometry already arises in two dimensions.
In fact, the spaces we will consider are twisted, there-dimensional tori, where a two-torus is non-trivially fibred over a circle in the third
direction \cite{Hull:2004in,Shelton:2005cf,Dabholkar:2005ve}. 
In the first example one starts with a flat three-torus with constant $H$-field. Here the $B$-field on the fibred two-torus is linear in the circle coordinate.
The corresponding monodromy transformations, called parabolic monodromies,  on the fibred two-torus are of infinite order, when transporting it 
along the base circle. 
As we will explain in more detail in the following,
there exists a chain of three T-duality transformations starting with the $H$  flux, leading to four different types of geometrical and non-geometrical fluxes:
\begin{equation}\label{eq:TdualityChain}
H_{abc} \stackrel{T_{a}}{\longrightarrow} f^{a}{}_{bc}
\stackrel{T_{b}}{\longrightarrow} Q_{c}{}^{ab}
\stackrel{T_{c}}{\longrightarrow} R^{abc}\, .
\end{equation}
Here $T_a$ denotes T-dualizing along direction $a$. The  $f$'s are called geometric fluxes, and they are given by the first derivatives of the vielbein and are related to the Levi--Civita spin connection and therefore to the curvature of the twisted torus (Nil-manifold). On the other hand, and the
$Q$'s and  the $R$'s  are non-geometric fluxes and their geometric meaning  will be discussed later in this paper.

As we will discuss in section three, 
for these kind of spaces, the  closed string geometry is non-commutative and also non-associative    
\cite{Blumenhagen:2010hj,Lust:2010iy,Blumenhagen:2011ph,Blumenhagen:2011yv,Condeescu:2012sp,Plauschinn:2012kd}.
Consider the commutator of the (dual) coordinates on the fibred two-torus. For non-vanishing fluxes $F$
they do not commute,
\begin{equation}\label{commclosed}  [ X^1, X^2 ]\simeq  F~ p^3\, ,   
  \end{equation}
Depending on the given duality frame,  the fluxes $F$ correspond to   either $F=H,f,Q$ or $R$-flux, respectively. The coordinates $X^1,X^2$
are either coordinates or dual coordinates on the fibred two-torus, again depending on the considered
duality frame. Finally,  $p^3$ is the momentum or the dual momentum in the third circle direction.
Second, the three-bracket between all three closed string coordinates or dual coordinates turns out to be non-vanishing:
\begin{equation}\label{asso}  [ X^1, X^2, X^3]\simeq  F\, .     
\end{equation}
This result shows that the $H,f,Q,R$-deformed closed string geometry exhibits not only non-commu\-ta\-ti\-ve but even genuine
non-associative structures.

Eqs.(\ref{commclosedb}) and (\ref{nonass}) together with $[X^k,p^k]=i$ now define a so-called {\sl twisted Poisson structure}. 
As discussed in \cite{Saemann:2012ex} (same proceedings), this algebra can  be nicely described by quantizing 2-plectic manifolds in loop space using groupoids.
Moreover, this structure also emerges
for the momenta of  point particles moving in the field of a magnetic monopole, where again the role of coordinates and momenta is exchanged
compared to the closed string geometry. We will argue that in closed string theory the emergence of this twisted Poisson structure is a stringy feature, related
to the fact that the closed string can move in non-geometric string backgrounds. Since T-duality
is relating left-right symmetric backgrounds  to left-right asymmetric closed string backgrounds, the non-commuative and non-associtive 
geometries in particular  arise for the coordinates of the left-right asymmetric spaces. In this sense, the $R$-flux background is left-right asymmetric in all
three world-sheet string coordinates.

For the case of constant $H$- resp. $R$-flux, i.e. for the parabolic case, the underlying closed string CFT and its correlation functions
were analyzed
in \cite{Blumenhagen:2011ph}.
The main result of this paper is that the algebra of periodic functions on these spaces, namely algebra of closed string (slighlty off-shell)  tachyon vertex operators,  
becomes non-associative. It is given in terms of a new non-associative 
$N$-product
\begin{eqnarray}
  & f_1(x)\, \tri_N\,  f_2(x)\, \tri_N \ldots \tri_N\,  f_N(x) \stackrel{\rm def}{=} \\
   &\exp\left[ {\textstyle {\pi^2\over 2}} F^{abc} \sum_{1\le i< j < k\le N}
     \!\!\!\!  \, 
      \partial^{x_i}_{a}\,\partial^{x_j}_{b} \partial^{x_k}_{c} \right]\, 
   f_1(x_1)\, f_2(x_2)\ldots
   f_N(x_N)\Bigr|_{x} \;,
\end{eqnarray}
which is the closed string generalization of the open string
non-commutative product eq.(\ref{Nbracketcon}).
This completely defines the new tri-product, which satisfies
the relation
\begin{eqnarray}
\label{thenicerel}
    f_1\,\tri_N\,  f_2  \,\tri_N\,  \ldots\,  \tri_N\, f_{N-1} \, \tri_N\, 1\,   \
   =  f_1\,\tri_{N-1}\,  \ldots\,  \tri_{N-1}\,  f_{N-1} \;.
\end{eqnarray}
Specializing above expression  to $N=3$ gives
\begin{eqnarray}
\label{threebracketcon}
   f_1(x)\,\tri\, f_2(x)\, \tri\, f_3(x) \stackrel{\rm def}{=} \exp\Bigl(
   {\textstyle {\pi^2\over 2}}\, \theta^{abc}\,
      \partial^{x_1}_{a}\,\partial^{x_2}_{b}\,\partial^{x_3}_{c} \Bigr)\, f_1(x_1)\, f_2(x_2)\,
   f_3(x_3)\Bigr|_{x} \;,
\end{eqnarray}

Besides the parabolic case, also twisted, there-dimensional tori with elliptic monodromy properties of the fibre two-torus can be studied. As we will discuss in the last section, 
here the elliptic monodromies are of finite order $N$, which means that the fibre two-torus comes back to itself when transported $N$-times
around the circle base. T-duality again maps a geometric background to a non-geometric $Q$ and $R$-flux background, which is non-commutative and
non-associative.\footnote{The closed string non-commutativity in relation with T-duality and non-geometric versus geometric string background was 
discussed for the first time in a model with elliptic $Z_4$-monodromy in \cite{Lust:2010iy}.} The corresponding fluxes are not anymore constant but given in terms
certain periodic functions of the base coordinates. Hence at first sight, it seems hard to argue that these spaces correspond to consistent conformal string backgrounds,
valid to all order in $\alpha'$. However it was shown \cite{Condeescu:2012sp} that at  a particular point in moduli spaces,  non-geometric backgrounds with elliptic monodrmies can be formulated as exact CFT's,
namely as symmetric or asymmetric freely acting orbifolds, depending on whether one is dealing with a geometric or non-geometric closed string background.
For these (a)symmetric orbifold spaces the full modular invariant partition function can be constructed, and the non-commuative behavior of the closed
string coordinates follows as an exact CFT result in the (a)symmetric orbifold limit.

\section{Twisted Poisson structures for point particles}

In this section we first  compare the non/commutative and non-associative closed string algebra, displayed in eqs.(\ref{commclosed}) and (\ref{asso}),
with the phase space structure of a particle moving a magnetic field. It turn out the the closed string case corresponds to a twisted Poisson structure
algebra, discussed in \cite{Jackiw:1984rd,Klimcik:2001vg,Alekseev:2004np,Kotov:2004wz}, of a point particle moving
in the spatially extended, non-constant magnetic field of a 
magnetic monopole. However going from the point particle to the closed string as a probe of the target space geometry, momentum and
position operators are exchanged.
A similar application applies for a point particle moving in a constant magnetic field as compared to the motion of an open string 
in the B- and F-field background described at the beginning: whereas for the point particle the position operators $X_i$ do commute, and the (mechanical) momentum operators do not commute,
in the open strings the position operators become non-commuting.

\subsection{Point particle in a constant magnetic field}

Let us first consider a charged point particle which is moving in a magnetic field $\vec B$.
The configuration space ${\cal M}$ is the tangent bundle of some manifold ${\cal Q}$:
\begin{equation}
{\cal M}=T^*{\cal Q}\, .
\end{equation}
The Langrange function of the particle (we sets its mass and its charge equal to one) with coordinate $x^i(t)$
is given by the follow expression
\begin{equation}
L={1\over 2}(p_i)^2={1\over 2}  (\dot x^i-A^i)^2\, ,
\end{equation}
where $\vec A(x)$ is the vectorpotential and $\vec B={\rm rot}~ \vec A$.
The canonical momenta $p_i={\partial L\over\partial \dot x^i}=\dot x^i-A^i$ satisfy the standard, canonical Poisson algebra:
\begin{equation}
\pi^{ij}=\lbrace x^i,x^j\rbrace=0\, ,\quad \lbrace p_i,p_j\rbrace=0\, ,\quad \lbrace x^i,p_j\rbrace=\delta_i^j\, .
\end{equation}
However this is not anymore true, when we consider the algebra of the mechanical momenta $\overline p^i=\dot x^i=p^i+A^i$.
Consider e.g. the simplest case of a 2-dimensional base manifold ${\cal Q}$ with a linear vector potential $\vec A$, i.e with a constant
magnetic field $\vec B$. Then one obtains after a short calculation the following Poission algebra:
\begin{equation}
\pi^{ij}=\lbrace x^i,x^j\rbrace=0\, ,\quad \pi_{ij}=\lbrace \overline p_i,\overline p_j\rbrace=\epsilon_{ijk} B^k\, ,\quad \lbrace x^i,p_j\rbrace=\delta_i^j\, .
\end{equation}
Hence, for  a particle moving in a constant magnetic field the mechanical momenta do not commute anymore, but their commutator is
given by the constant magnetic field. As its is well known, this form of non-commtatative algebra occurs for the open string coordinates on
$D2$-branes moving in the
background of a constant magnetic field. Note that comparing point particles
with the open string, the role of momenta and positions are essentially exchanged.

\subsection{Point particle in the field of a magnetic monopole}

Now we consider a base manifold ${\cal Q}$, which is at least 3-dimensional.
The $B$-field is non anymore assumed to be constant and can be even topologically non-trivial:
\begin{equation}
\vec B\in H^2({\cal Q})\, 
\end{equation}
Therefore we allow the possibility that $B$ is non-closed:
\begin{equation}
H=dB=\star\rho_{magn}\, .
\end{equation}
In three dimensions, $\rho_{magn}$ is the magnetic charge density of a magnetic monopole, which is smeared out over space.
Let us again assume the simplest case that the 3-form $H$ is constant over space, i.e. the magnetic field $\vec B$ is a spatially linear 2-form.
This implies the following new Poisson algebra:
\begin{equation}
\pi^{ij}=\lbrace x^i,x^j\rbrace=0\, ,\quad \bar\pi_{ij}=\lbrace \overline p_i,\overline p_j\rbrace=H_{ijk}x^k\, ,\quad \lbrace x^i,p_j\rbrace=\delta_i^j\, .
\end{equation}
In addition, using the still valid canonical relation $\lbrace x^i,p_j\rbrace=\delta_i^j$, one discovers that this algebra is even non-associative.
This means that Jacobi-identity among the mechanical momenta is not anymore satisfied, i.e. the double Poisson-bracket plus
its permutations is non-vanishing:
\begin{equation}
\bar \pi_{ijk}=\lbrace \lbrace \overline p_i,\overline p_j\rbrace,\overline p_k\rbrace+{\rm perm.}=H_{ijk}
\end{equation}
This algebra is called {\sl twisted Poisson structure}, it is non-commutative as well as non-associative with respect to the mechanical momentum
variables. In this simplest case, the right hand side of the non-associative 3-bracket is a constant, namely just the exterior derivate of the
magnetic field. As we will see in the following, the analogous twisted Poisson structure structure is present for closed strings coupled to a non-vanishing $H$-field background.
However, the difference between point particles and the closed string is that coordinates and momenta will exchange their roles.

\section{Twisted Poisson structures from closed strings on non-geometric backgrounds}

\subsection{The T-dual chain of geometric and non-geometric flux backgrounds}

\subsubsection{Geometric flux backgrounds}

In the simplest case for constant background metric $g$ and $b$-field, $SO(d,d)$ T-duality transformations are just acting as 
automorphism on the moduli space of the string background space.
However for non-constant background fields, T-duality transformations in general not only change the geometry of the associated
background, but can rather lead to a topology change. This is in particular true for backgrounds with non-vanishing NS
$H$-flux, i.e. where the $b$-field is non-constant. So for concreteness, let us consider the compactification on a three-dimensional
torus $T^3$, which can be viewed as a ${\cal F}=T^{n}$ fibration  $(n=0,1,2,3$) over a (3-n)-dimensional base
${\cal B}=T^{3-n}$. In the following we will consider four different cases.

\vskip0.2cm
\noindent {\sl (Frame A) $n=0$: The $H$-flux space:}

\vskip0.2cm
\noindent
Here there are $N$ units of constant, isotropic $H_{abc}$-flux on $T^3$, i.e. $H=Ndx \wedge dy\wedge dz$. Note, however, that
the associated, non-constant $b$-field breaks the isotropy of the background, namely when writing it in a particular gauge as $b=Nz\,dx\wedge dy$.
In the following we will perform $T$-duality transformations over all directions of the $n$-dimensional $T^{n}$ fibre torus, and hence we discuss three additional  background spaces
  \cite{Shelton:2005cf,Dabholkar:2005ve,Hull:2006va}:

\vskip0.2cm
\noindent {\sl (Frame B) $n=1$: The $f$-flux space:}

\vskip0.2cm
\noindent
Performing a T-duality transformation on the one-dimensional circle fibre ${\cal F}=T^{1}_x$ in $x$-direction, one obtains the Heisenberg nilmanifold,
which is a twisted torus without $b$-field. It is topologically distinct from $T^3$, since it has $H_1(Z)=Z\times Z\times Z_N$.
Its Levi-Civita connection can be related to the structure constants $f^a_{bc}\sim N$ of the Heisenberg group, where the 
\begin{equation}
f^a_{bc}=-2e_{[b}^me_{c]}^n\partial_me_n^a
\end{equation}
are called geometrical fluxes.
The effective action for the geometric string backgrounds of frames A and B is given in terms  of  the well-known low-energy supergravity action, its bosonic part is the standard  Neveu--Schwarz (NSNS) Lagrangian:
\begin{equation}
{\cal L}= e^{-2\phi} \sqrt{|g|} \left(\R + 4(\partial \phi)^2 - \frac{1}{12} H_{ijk} H^{ijk} \right) . \label{eq:Lns}
\end{equation}

\subsubsection{Non-geometric flux backgrounds and doubled field theory}

\vskip0.2cm
\noindent {\sl (Frame C) $n=2$: The $Q$-flux space:}

\vskip0.2cm
\noindent
Performing a T-duality transformation in  the $x,y$-directions on the two-dimensional torus fibre ${\cal F}=T^{2}_{x,y}$, one obtains the so-called $Q$-space.
This new background is again a $T^2$-fibration. But the corresponding metric and $b$-field are only locally defined, but not anymore globally.
Hence this space is one of the simplest examples of a non-geometric string background that is not anymore a Riemannian
manifold. The reason for the failure of being a Riemannian manifold is given by the observation
that the fibre ${\cal F}$ has to be glued together by a T-duality transformation when transporting it once around the base ${\cal B}$, and not by a standard diffeomorphism.
Concretely for our example, the K\"ahler structure $\rho=b+iVol(T^2)$  of ${\cal F}$ has the  (parabolic) monodromy property
\begin{equation}\label{rhotrans}
\rho(z+2\pi)={\rho(z)\over 1+2\pi N\rho(z)},
\end{equation}
which is just an infinite order element of the $T$-duality group $SO(2,2,Z)$.  Here, the associated $Q$-flux shows up in eq.(\ref{rhotrans}) as
the discrete parameter, which determines the non-trivial modular transformation of $\rho(z)$.

The geometric role of the non-geometrtric fluxes can be nice described via the frame work of doubled field theory (DFT).
DFT was introduced in \cite{Hull:2009mi,Hull:2009zb,Hohm:2010jy,Hohm:2010pp}. 
In this theory, T-duality is turned into a manifest symmetry by doubling the coordinates at the level of the effective space-time action for string theory. T-duality relates momentum and winding modes of a closed string moving on a torus $T^D$ via the T-duality group $O(D,D)$.  When the coordinates are doubled, this duality symmetry can be made manifest. 
Thus, in DFT every conventional coordinate $x^i$, associated to momentum modes, is complemented by a dual coordinate $\tilde{x}_i$, associated to winding modes. The coordinates combine into a fundamental $O(D,D)$ vector $X^{M}=(\tilde{x}_i,x^{i})$.

Now consider the following field redefinition of the metric $g$ and the $b$-field:
\begin{equation}
\label{eq:relation}
(\tilde g^{-1} + \beta)^{-1} \equiv \tilde{\cal E}^{-1} = {\cal E} =  g + b \ ,
\end{equation}
where we have introduced
\begin{equation}
\tilde{\cal E}^{ij} = \tilde g^{ij} + \beta^{ij} \ ;
\end{equation}
here $\beta^{ij}$ is a bi-vector.
We can  also redefine $\phi$:
\begin{equation} \label{eq:dilintro}
\sqrt{|g|} e^{-2\phi} = e^{-2d} = \sqrt{|\tilde g|} e^{-2\tilde \phi} \ .
\end{equation}
The redefinition (\ref{eq:relation}) has the form of an overall T-duality, here in all three toroidal directions.
Following the recent work on the effective action of non-geometric fluxes \cite{Andriot:2011uh,Andriot:2012wx, Patalong:2012np,Larfors:2012zz,Andriot:2012an}, the $Q$-flux is now defined as  
 \begin{equation}
\label{eq:qflux}
  Q_{m}{}^{nk} = \partial_{m}\beta^{nk}\;.
 \end{equation}
 Note that, being a partial derivative of a bi-vector, $Q$ is not a tensor. However, as shown in \cite{Andriot:2012wx,Andriot:2012an},
 the proper geometrical interpretation of $Q$ is playing the role of a connection,
 which allows us to construct a 
derivative for the dual $\tilde{x}$ coordinates that is covariant  \textit{with respect to the $x$ diffeomorphisms}. 
In addition one can show that for certain non-geometric situations, where the metric $g$ and the $b$-field are only locally but not globally 
defined,  the $Q$-flux is nevertheless a globally well-defined object.  In particular for the chain of T-duality transformations with constant flux, it turns
out that $\beta^{ij}$ is linear in the circle coordinate $z$, $\beta^{xy}=Nz$, and hence the $Q$-flux is simply given by $N$. i.e. $Q_{z}{}^{xy}=N$.

\vskip0.2cm
\noindent {\sl (Frame D) $n=3$: The $R$-flux space:}

\vskip0.2cm
\noindent
Finally one can consider a T-duality transformation along the entire three-dimensional  torus, being seen as a fibration with fibre ${\cal F}=T^{3}_{x,y,z}$ over a point. 
However this T-duality is much more problematic, since the $z$ direction is not any longer a Killing isometry of the background space. Hence the standard
Buscher rules cannot be applied, and T-duality acts like a formal transformation in CFT, that leads to a "space" that is left-right asymmetric in all its three directions.
This new background is not known explicitly, in fact it is not even locally a Riemannian manifold, describable by a metric $g$ and a $b$-field.
However, as shown in    \cite{Andriot:2012wx,Andriot:2012an}, the associated $R^{abc}$-flux can be again constructed in a geometric manner when using the formalism of doubled field theory.
The reason, why this will be possible, is that in DFT T-duality transformations can be performed regardless if one is dealing with a Killing isometry or not.
The $R$-flux can be viewed as the tensor, which is completely T-dual in all directions  to the standard $H$-flux. Its precise definition is given by the following equation, 
\be\label{eq:rflux}
  R^{ijk}  =  3\tilde{D}^{[i}\beta^{jk]}  \;,
 \ee
where we introduce the derivative operator
\be\label{eq:Dtilde}
  \tilde{D}^{i} \ \equiv \ \tilde{\partial}^i -\beta^{ij}\partial_{j}  \; .
 \ee
This $R$ flux is a tensor and represents the covariant field strength of $\beta$.
In the supergravity limit the DFT fields are taken to be independent of the dual  coordinates, i.e. one sets $\tilde{\partial}^i = 0$ in the action.
Here the $R$ flux term is given as
\be
R|_{\tilde{\partial}=0}^{ijk}  = 3 \beta^{p[i}\partial_{p}\beta^{jk]} \;.
 \ee
 This expression is still covariant.
Moreover, in DFT we may equally well solve the strong constraint by setting the conventional derivatives to zero, $\partial_i=0$, 
keeping the winding derivatives $\tilde{\partial}^i$. This corresponds to a T-duality inversion in all directions. 
The $R$ flux now reads 
\be
 R^{ijk} \ = \ 3\tilde{\partial}^{[i}\beta^{jk]}\;.
\ee
Note that  the $R$ flux together with the $Q$-flux satisfies the following Bianchi identity  \cite{Andriot:2012wx,Andriot:2012an}\footnote{This Bianchi identity was also derived by a different method in \cite{Blumenhagen:2012ma}.} 
  \be\label{RfluxBianchi}
   \tilde{\nabla}^{[i}R^{jkl]} \ = \ 0\;, 
  \ee
which reads explicitly 
  \be
  4\,\tilde{\partial}^{[i}R^{jkl]}+4\,\beta^{p[i}\partial_{p}R^{jkl]}+6\,Q_{p}{}^{[ij} R^{kl]p} \ = \ 0\;. 
 \ee

Via DFT or via field redefinitions we can also  formulate an effective action for non-geometric string backgrounds. This new effective action is particularly useful
in case the standard $b$-field and the standard metric $g$ are not well-defined quantities, as it is generically the case for non-geometric string background spaces.
Specifically, 
the DFT action in terms of the new fields takes the following schematic form:
\begin{eqnarray}
\label{NewDFTaction}
S_{\rm DFT}(\tilde g,\b,\tilde\phi ) &= & \int   dxd\tilde{x}\,\sqrt{|\tg|}\,e^{-2\tilde{\phi}}\Big[{\cal R} (\tilde g,\partial)
+{\cal R} (\tilde g^{-1},\tilde{\partial})\\ \nonumber
&-&\frac{1}{4}Q^2 -\frac{1}{12}R^{ijk}R_{ijk} 
  +4\Big( (\partial\tilde{\phi})^2
  + (\tilde{\partial}\tilde{\phi})^2 \Big) +\dots\Big] 
\end{eqnarray}
There are two Einstein-Hilbert terms: one based on the conventional derivative $\partial_{i}$, and one based
on the winding derivatives $\tilde{\partial}^{i}$, where the inverse metric $\tilde{g}^{ij}$ plays the role of the 
usual metric, and so works consistently with the lower indices of the winding coordinates $\tx_{i}$. 
Even though the first Einstein-Hilbert term is manifestly invariant under $x$ diffeomorphisms $x^{i}\rightarrow x^{i}-\xi^{i}(x)$,
and the second Einstein-Hilbert term is manifestly invariant under $\tilde{x}$ diffeomorphisms $\tilde{x}_{i}\rightarrow \tilde{x}_{i}-\tilde{\xi}_{i}(\tilde{x})$, 
the invariance of the full action as written in (\ref{NewDFTaction}) is not manifest for either of them. 
The reason is that in the full DFT the parameters $\xi^{i}$ and $\tilde{\xi}_{i}$ can a priori depend both on $x$ and $\tilde{x}$. Moreover, 
as mentioned above, $Q$ is not a tensor and therefore the $Q^{2}$ term is not separately diffeomorphism invariant. 
As we have shown,  in our formalism precisely half of the gauge symmetries can be made manifest, here the 
diffeomorphisms parameterized by $\xi^{i}$, by introducing a novel tensor calculus. The $Q$ can then be interpreted as the antisymmetric part of the 
`dual' connection coefficients, so that the $Q^{2}$ term is just part of an extended dual Einstein-Hilbert term.

In summary,  we finally have introduced the 
following chain of three T-duality transformations relating four different types of geometrical and non-geometrical spaces:
\begin{equation}
H_{abc} \stackrel{T_{x}}{\longrightarrow} f^{a}{}_{bc}
\stackrel{T_{y}}{\longrightarrow} Q_{c}{}^{ab}
\stackrel{T_{z}}{\longrightarrow} R^{abc}\, .
\end{equation}
In the next section we will discuss that depending on the considered duality frame, the phase space of the theory is described by a non-commutative and non-associative
twisted Poisson structure, namely either for the coordinates or for the dual coordinates.

\subsection{Non-commutative and non-associative geometry from fluxes}

In the following we will relate  the background fluxes  to the non-commutative and non-associate geometry structure,
which is present in non-geometric string backgrounds. Concretely, a non-commutative algebra for the string coordinates emerges for the $Q$-flux spaces, and the $R$-flux backgrounds even lead to
a non-associative algebra structure.\footnote{These non-commutative and non-associatives structures also appeared in more mathematics oriented
literature  \cite{Bouwknegt:2000qt,Bouwknegt:2004ap,Mathai:2004qq,Mathai:2004qc,Grange:2006es,Grange:2007bp,Brodzki:2007hg}, where vibrations and twisted K-theory are applied to characterize
these kind of non-geometric backgrounds with D-branes and $B$-fields.}
It is the non-geometric flux  $R=\tilde \partial \beta$, which corresponds to parameter that
controls 
the violation of the Jacobi identity, i.e. to the deformation parameter of the non-associative algebra of the $R$-flux backgrounds.

Let us therefore try to relate the geometrical objects $\beta^{mn}$, $Q_m{}^{nl}$ and $R^{mnl}$ of DFT to the deformation parameters of the associated non-commutative respective non-associative algebras.

\subsubsection{Open strings}

One first relevant observation in this context is that these objects are closely related to the non-commutative open string geometry on these spaces. 
Consider frame (A) with a $b$-field on the two-dimensional fibre torus ${\cal F}=T^{2}_{x,y}$:
\begin{equation}
b=b(z)\,dx\wedge dy
\end{equation}
The corresponding $H$-field
is given as 
\begin{equation}
H={\partial b(z)\over \partial z}\,dx \wedge dy\wedge dz
\end{equation}
One can easily convince oneself that the open string metric $g_{\rm open}$ and the open string non-commuta\-ti\-vi\-ty deformation $\theta$ (see \cite{Seiberg:1999vs}) parameter correspond to the dual metric $\tilde g$ and the bi-vector $\beta$. 
Note that $\tilde g_{\rm open}$ is just the open string metric in the limit where gravity is decoupled from the D2-brane world volume. Concretely, for D2-branes that are wrapped around the torus fibre of the $H$ flux space,
one obtains the following equal-time commutator for the open string coordinates (at the location $\sigma = 0,\pi$ of the D2-brane):
\begin{equation}\label{commopen}  [ X^m(\tau), X^n(\tau) ]_{\rm open}=     \beta^{mn} \, . 
\end{equation}
This defines a so-called Poisson structure in analogy to the momentum algebra of a point particle
moving in a (constant) magnetic field.

\subsubsection{Closed strings}

We will start the discussion in the duality frame C, namely we will describe
the non-commuta\-ti\-ve geometry of closed strings moving in the non-geometric $Q$ flux background.
A first guess could be that the non-commutativity is again directly related to the bi-vector $\beta$, leading to the same algebra
(\ref{commopen}) as for the open strings.
However, this will not be quite correct: as discussed in \cite{Lust:2010iy,Condeescu:2012sp} (see also next section), only an extended closed string which is wrapped $\tilde p^k$ times around the base of the fibration is sensitive to the global ill-definedness of the two-dimensional fibre torus. As a result the fibre geometry becomes non-commutative with non-commutativity deformation parameter given in terms of  the winding number $\tilde p^k$: 
\begin{equation}\label{commclosed} 
 [ X^m(\tau,\sigma), X^n(\tau,\sigma) ]_{\rm closed}\sim   \epsilon_k{}^{mn}\tilde p^k \, .
\end{equation}
In view of this result, we propose the following integral relation between the non-geometric $Q$ flux and the closed string non-commutativity \cite{Andriot:2012an}:
\begin{equation}\label{commcloseda}
 [ X^m(\tau,\sigma), X^n(\tau,\sigma) ]_{\rm closed}=  \oint_{C_k}Q_k{}^{mn}(X)~ dX^k \ ,
\end{equation}
where $C_k$ is a non-trivial homology base cycle, around which the closed string is wrapped $\tilde p^k$ times. In the case of constant flux $Q_k{}^{mn}=Q\epsilon_k{}^{mn}$ and $C_k=S^1$ one gets
\begin{eqnarray}\label{commclosedb}
 [ X^m(\tau,\sigma), X^n(\tau,\sigma) ]_{\rm closed}= \oint_{C_k}Q_k{}^{mn} dX^k \;
 = \; 2\pi Q  \epsilon_k{}^{mn}\tilde p^k    \, ,
\end{eqnarray}
in agreement with (\ref{commclosed}). 
So in more general terms, the right hand side of the commutator is given in terms of the {\sl Wilson line operator} of the gauge connection $Q$. Although the $Q$-flux not being a covariant object,
it is conceivable that the Wilson line operator is indeed a covariant object under coordinate transformations.

Next we turn to the duality frame D, and we
discuss the $R$ flux background obtained by a T-duality transformation, $X^k\leftrightarrow \tilde X_k$, in the $k^{\rm th}$ direction from the previous case.
The corresponding closed string background becomes non-associative, as discussed in \cite{Blumenhagen:2010hj} in the context of the $SU(2)$
Wess--Zumino--Witten model, and investigated  in \cite{Blumenhagen:2011ph}
by  the computation of conformal field theory  amplitudes in the chain of T-dual $H,f,Q,R$-backgrounds
leading to a non-associative algebra of closed string vertex operators.
So it is quite natural to conjecture that the non-geometric flux  $R$  corresponds to the parameter that
controls 
the violation of the Jacobi identity, i.e. to the deformation parameter of the non-associative algebra of the $R$ flux backgrounds:
\begin{equation}\label{nonass} 
 [[ X^m(\tau,\sigma), X^n(\tau,\sigma)],X^k(\tau,\sigma)]_{\rm closed} +{\rm perm.} =     R^{mnk}\ .
 \end{equation}
Note that this non-associativity relation can be at least formally derived from the commutator (\ref{commclosedb}) by using the Heisenberg commutation relation
$[X^k,p^k]=i$
in the $k^{\rm th}$ direction.

By T-duality 
 we can even include the geometrical spaces (Frames A and B) in our discussion. For the $H$-flux background as well as for the geometrical
 $f$-flux the coordinates are still commutative and associative. For these backgrounds, the non-commutativity  however shows up when including the dual
 coordinates of these spaces. Without going into the details of all possible T-duality transformations, we summarize the results
  in  table 1, where
 the various commutation relations between coordinates $X^a$ and their duals  $\tilde X^a$  are listed.
\begin{table}[t]
\centering
\renewcommand{\arraystretch}{1.3}
\tabcolsep10pt
\begin{tabular}{|c||c|c|}
\hline
Flux  & Commutators & Three-brackets \\  \hline\hline
Frame A: $H$-flux & $[\tilde X,\tilde Y]\simeq H \tilde p_z$ & $[\tilde X,\tilde Y, \tilde Z]\simeq H$\\
Frame B: $f$-flux & $[X,\tilde Y]\simeq  f\tilde p_z$ & $[ X,\tilde Y, \tilde Z]\simeq f$\\
Frame C: $Q$-flux & $[ X,Y]\simeq Q \tilde  p_z$ & $[X,Y, \tilde Z]\simeq Q$ \\
Frame D: $R$-flux & $[ X,Y]\simeq R p_z$ & $[ X,Y, Z]\simeq R$ \\ \hline
\end{tabular}
\caption{\small  Non-vanishing commutators and three-brackets in the four flux backgrounds.
\label{tablemomwind2} } 
\end{table}

\section{(Non)-geometric backgrounds with elliptic monodromies and their (a)symmetric orbifold CFT's}

In this chapter we will derive the algebraic structures, discussed in the previous section, within a concrete class of string backgrounds, namely
non-geometric backgrounds with elliptic monodromies. We will first show the derivation of the non-commutative closed string geometry using a twisted
mode expansion for the coordinates, that reflects the monodromy properties of the non-geometric string background \cite{Lust:2010iy}. After that we will provide an exact non-commutative
asymmetric orbifold CFT, that is valid at the orbifold point of the non-geometric background \cite{Condeescu:2012sp}.

\subsection{Non-commutativity from elliptic monodromies and twisted closed string mode expansion}

So let us start to discuss torus fibrations with elliptic monodromies and their non-commutativity. Here, we are dealing with a chain of two T-dualities leading to
three different dualities frames\footnote{For this class of models there is no T-duality transformation that brings us to a background with only non-trivial $H$-flux (frame A).}

\vskip0.2cm
\noindent {\sl (Frame B): The $f$-flux space:}

\vskip0.2cm
\noindent
We consider again a twisted three-torus $T^3$, being a torus $T^2$ fibered over $S^1$ with coordinates $X^1, X^2$ on the fiber and $\mathbb{X}$ on the base.  The metric of the total space is of the of the following form:
\begin{equation}
ds^2=\frac{1}{ \tau_2(\mathbb{X})}\left|dX^1+\tau(\mathbb{X})dX^2\right|^2+d\mathbb{X}^2~
\end{equation}
with the complex structure $\tau=\tau_1+i\tau_2$ of the fiber being, in general, a non-trivial function of the base space coordinate $\mathbb{X}$. An important class of torus fibrations is given by elliptic monodromies, which act on the toroidal coordinates as rotations. For example, consider the monodromy corresponding to a $\mathbb{Z}_4\subset O(2;\mathbb{Z})$ rotation:
\begin{equation}\label{z4}
\left(
  \begin{array}{c}
    X^1 \cr
    X^2 
  \end{array}
\right) \rightarrow \left(
  \begin{array}{c}
    X^2 \cr
    -X^1 
  \end{array}
\right)~,
\end{equation}
resulting in an $SL(2;\mathbb{Z})$-transformation of the complex structure of the fiber:
\begin{equation}
\tau(\mathbb{X})\rightarrow -1/\tau(\mathbb{X})~.
\end{equation}
As we will discuss in the next subsection, for $\tau(\mathbb{X})=i$, the complex structure is a fixed point of the above transformation. At this point in moduli space, the fibered torus admits an exact CFT description in terms of a freely-acting $\mathbb{Z}_4$-orbifold, corresponding to the minimum of the Scherk-Schwarz potential for the complex structure.

One can diagonalize the $\mathbb{Z}_4$ rotation (\ref{z4}) by introducing complex coordinates $Z=\frac{1}{\sqrt{2}}(X^1+iX^2)$ and, hence, obtain the following twisted boundary conditions:
\begin{equation}\label{IntroBC}
Z(\tau,\sigma+2\pi)=e^{2\pi i \theta}Z(\tau,\sigma)~,
\end{equation}
with the angle $\theta$ depending on the winding number (dual momentum) $\tilde p^{\mathbb{X}}$ in the $S^1$-direction:
\begin{equation}
\theta=-f\tilde p^{\mathbb{X}} \quad, \quad {\rm with\ }~~ f\in\frac{1}{4}+\mathbb{Z}~.
\end{equation}
Even though the complex structure $\tau(\mathbb{X})$  is a non-trivial function of the base coordinate $\mathbb{X}$ and the $\sigma$-model is, in general, non-linear, one could still  write down a mode expansion for the fiber coordinates subject to the twisted boundary conditions (\ref{IntroBC}), with the understanding that the result is only a lowest-order approximation in the flux (and in $\alpha'$). Introducing the usual left- and right-moving coordinates $Z_{L,R}$, one obtains:
\begin{eqnarray}
Z_L(\tau+\sigma)&=&\frac{i}{\sqrt{2}}\sum_{k\in\mathbb{Z}}\frac{\alpha_{k-\theta}}{k-\theta}e^{-i(k-\theta)(\tau+\sigma)}~,\nonumber\\
Z_R(\tau-\sigma)&=&\frac{i}{\sqrt{2}}\sum_{k\in\mathbb{Z}}\frac{\tilde\alpha_{k+\theta}}{k+\theta}e^{-i(k+\theta)(\tau-\sigma)}~.
\end{eqnarray}
Similar expansions hold for the complex conjugates, $\bar Z_{L,R}$. The usual quantization procedure then leads to the familiar bosonic oscillator algebra for the complex-conjugate Fourier modes:
\begin{equation}
[\alpha_{k-\theta},\bar\alpha_{\ell-\theta}]=(k-\theta)\delta_{k,\ell}~.
\end{equation}
Explicit calculation then yields the equal-$\tau$ commutation relations for the coordinates\footnote{Strictly speaking, the commutation algebra should include the metric factor $G^{z\bar z}$, but this is immaterial for our present discussion and we will simply suppress it.}:
\begin{equation}\label{IntroSeries}
[Z_L(\tau,\sigma),\bar Z_L(\tau,\sigma')]=\frac{1}{2}\sum_{k\in\mathbb{Z}}\frac{e^{-i(k-\theta)(\sigma-\sigma')}}{k-\theta}\equiv \frac{1}{2}\Theta(\sigma-\sigma',\tilde p^{\mathbb{X}})~.
\end{equation}
For the right-moving coordinates, a similar calculation yields  the same function $\Theta$, but with the opposite sign:
\begin{equation}
[Z_R(\tau,\sigma),\bar Z_R(\tau,\sigma')]=-\frac{1}{2}\Theta(\sigma-\sigma',\tilde p^{\mathbb{X}})~.
\end{equation}
In order to obtain a local result, one should carefully investigate the limit $\sigma\rightarrow \sigma'$. Since in this limit, the series representation (\ref{IntroSeries}) is naively divergent, it has to be defined through analytic continuation.
To this end, upon introducing the complex variable $z=e^{i(\sigma'-\sigma)}$, the function $\Theta$ can be neatly represented in terms of hypergeometric functions:
\begin{equation}\label{hypergeometric}
\Theta(\sigma-\sigma',\tilde p^{\mathbb{X}})=-\frac{z^{-\theta}}{\theta}\Big[ {}_2F_1(1,-\theta;1-\theta;z)+ {}_2F_1(1,\theta;1+\theta;z^{-1})-1\Big]~,
\end{equation}
which can then be analytically continued to $z\rightarrow 1$. The final result for $\Theta$ is given in terms of elementary functions:
\begin{equation}
\Theta(\tilde p^{\mathbb{X}}) \equiv \left\{ \begin{array}{c l}
			-\pi\cot (\pi\theta) & , \quad \theta\notin \mathbb{Z} \cr
			0 & , \quad \theta \in \mathbb{Z} \cr
				\end{array}\right. ~,
\end{equation}
modulo the discontinuity at $\theta\in\mathbb{Z}$, arising from the subtraction of the zero mode in (\ref{IntroSeries}).

As expected for a geometric background, adding together the left- and right-moving coordinates, the twisted torus leads to commutative coordinates:
\begin{equation}
[X^1(\tau,\sigma),X^2(\tau,\sigma)]=0~,
\end{equation}
where we have expressed the commutation relations in terms of the original coordinates $X^1,X^2$ of the torus fiber.

\vskip0.2cm
\noindent {\sl (Frame C): The $Q$-flux space:}
\vskip0.2cm
\noindent
Performing a T-duality in the $X^1$-direction, one is lead to a non-commutative coordinate algebra. Indeed, introducing the dual coordinate $\tilde X^1=X^1_L-X_R^1$, one obtains:
\begin{equation}
[\tilde X^1(\tau,\sigma),X^2(\tau,\sigma)]=[X_L^1(\tau,\sigma),X_L^2(\tau,\sigma)]-[X_R^1(\tau,\sigma),X_R^2(\tau,\sigma)]=i\,\Theta(\tilde p^{\mathbb{X}})~.
\end{equation}

The same commutation algebra arises by imposing from the very beginning the following asymmetric boundary conditions:
\begin{eqnarray}
Z_L(\tau,\sigma+2\pi)&=&e^{2\pi i\theta}Z_L(\tau,\sigma)~,\nonumber\\
Z_R(\tau,\sigma+2\pi)&=&e^{-2\pi i\theta}Z_R(\tau,\sigma)~,
\label{asymmetric}
\end{eqnarray}
which are induced by the following monodromy transformation now acting on the K\"abler parameter $\rho(\mathbb{X})$ of the two-torus in the following way:
\begin{equation}
\rho(\mathbb{X})\rightarrow -1/\rho(\mathbb{X})~.
\end{equation}
Since this transformation is a non-trivial element of the T-duality group, the background is non-geometrical, i.e. a $Q$-flux space.
Moreover, these transformations are highly reminiscent of an asymmetric orbifold. The first line in eq. (\ref{asymmetric}) corresponds to the $\mathbb{Z}_4$-action:
\begin{equation}
X_L^1\rightarrow X_L^2 \quad, \quad X_L^2\rightarrow -X_L^1~,
\end{equation}
whereas the second line now gives for the right-moving sector:
\begin{equation}
X_R^1\rightarrow -X_R^2 \quad, \quad X_R^2\rightarrow X_R^1~.
\end{equation}
These asymmetric rotations define the following monodromy element $g$ of the $O(2,2;\mathbb{Z})$ T-duality group:
\begin{equation}
g =\left(
            \begin{array}{cccc}
              0 & 0 & 0 & 1 \cr
              0 & 0 & -1 & 0 \cr
              0 & 1 & 0 & 0 \cr
              -1 & 0 & 0 & 0 \cr
            \end{array}
          \right)~.
\end{equation}
Notice that $g$ is no longer an element of the geometric subgroup $GL(2;\mathbb{Z})$.

Now using the corresponding asymmetric mode expansion and adding left- and right-moving coordinates, one obtains for the commutator of the torus coordinates:\footnote{More general asymmetric rotations are possible for other choices of $\theta_L$ and $\theta_R$, provided they are compatible with modular invariance. In this case, the algebra of coordinates becomes $[X^1,X^2]=\frac{i}{2}\left\{\Theta(\theta_L)-\Theta(\theta_R)\right\}$.}   
\begin{equation}
[ X^1(\tau,\sigma),X^2(\tau,\sigma)]=i\,\Theta(\tilde p^{\mathbb{X}})~.
\end{equation}

The arguments presented above, hence, support the following conclusion. In stringy non-geometric compactifications, non-commutativity for the closed string coordinates can arise whenever the monodromy matrix $g\in O(N,N;\mathbb{Z})$ is not an element of the geometric subgroup $GL(N;\mathbb{Z})$. Furthermore, for special points in the moduli space of the theory, an exact CFT description in terms of freely-acting asymmetric orbifolds may exist for such non-geometric models. Of course, the construction of asymmetric orbifolds is not automatic. In general, modular invariance severely constrains the space of consistent asymmetric orbifold vacua. In the next subsection we will present explicit constructions of freely-acting asymmetric toroidal orbifolds and show  how non-commutativity arises in these setups. In particular, our results will be exact to all orders in $\alpha'$.

\vskip0.2cm
\noindent

\vskip0.2cm
\noindent {\sl (Frame D): The $R$-flux space:}

\vskip0.2cm
\noindent
The R-flux space is reached by performing a T-duality transformation in the ${\mathbb{X}}$-direction. Although we cannot write down a local metric for this background, one can perform
the T-duality transformation on the level of the non-commutative algebra by simply replacing the winding number, i.e. the dual momentum number $\tilde p^{\mathbb{X}}$, by
its T-dual  KK momentum number $p^{\mathbb{X}}$. Then the algebra of two-torus coordinates becomes:
\begin{equation}
[ X^1(\tau,\sigma),X^2(\tau,\sigma)]=i\,\Theta( p^{\mathbb{X}})\quad {\rm with}\quad \Theta( p^{\mathbb{X}})=\pi\cot (\pi f p^{\mathbb{X}})\, .
\end{equation}
Using the the Heisenberg commutation relation
$[\mathbb{X},p^{\mathbb{X}}]=i$ and linearizing the $\cot$-function one finally obtains the non-associative structure
\begin{equation}\label{nonass} 
 [[ X^1(\tau,\sigma), X^2(\tau,\sigma)],\mathbb{X}(\tau,\sigma)]]=     f\pi^2\ .
 \end{equation}

\subsection{Non-commutativity in asymmetric orbifold CFT's}

The fully consistent study of non-commutative effects in string theory necessitates a treatment that is exact to all orders in $\alpha'$. 
A natural candidate is provided by a class of freely-acting asymmetric $\mathbb{Z}_N$-orbifolds. The advantage of considering such backgrounds is that the corresponding worldsheet CFT is locally free, allowing us to obtain exact mode expansions for the internal coordinates and their commutators, $[X^I,X^J]$. 
We will start with the construction of a simple class of freely-acting asymmetric orbifold models. To illustrate the consistency of the models, we will explicitly display their modular invariant partition function.  These models should be considered as special solvable points in the moduli space of more general, non-geometric $Q$-flux backgrounds. Hence, these results will be exact to all orders in $\alpha'$ and to all orders in the (quantized) value of the flux\footnote{Contrary to the case of non-freely acting orbifolds, in which the internal space is singular at fixed points, the orbifolds we construct are free of such singularities, due to their freely-acting nature.}.

The constructions \cite{Condeescu:2012sp} we present will be Type II (freely-acting) asymmetric orbifold models with $4\leq\mathcal{N}_4< 8$ spacetime supersymmetry, compactified on $(S^1\times T^5)/\mathbb{Z}_N$. The action of the $\mathbb{Z}_N$ on $T^5$ will be specified below. Let us also note that the restriction to Type II theories is only a convenient choice, as our results can be extended to Heterotic theories in a straightforward fashion.

We will start our discussion by considering an $N$-dimensional torus $T^N\subseteq T^5$, that is locally factorized from an $S^1$-circle of radius $R$. We will then proceed by defining the action of $\mathbb{Z}_N$ on this manifold and derive various consistency conditions, such that the theory is a well-defined asymmetric orbifold. In fact, in what concerns the $T^N$, we will restrict our attention to asymmetric orbifolds where the $\mathbb{Z}_N$ acts non-trivially only on the left-moving degrees of freedom. To this end, the discussion in this section refers to the left-moving worldsheet fields. It is then easy to extend the construction to the right-moving sector in order to define symmetric orbifolds as well.

Let us take the $N$-dimensional torus $T^N$ to be parametrized by coordinates $X^I$, with $I=1,\ldots, N\leq 5$ and, further denote the coordinate associated to the $S^1$-circle by $\mathbb{X}$. We will restrict ourselves to the case where the orbifold acts as a permutation $P$ (including a possible `reflection') only on the left-moving coordinates of $T^N$, while leaving the right-movers invariant. Furthermore, in order to eliminate fixed points and generate a free action, we will couple this to a shift both in the momenta and windings of the $S^1$-direction in the orbifold basis. The orbifold element can be expressed, hence, as:
\begin{equation}\label{GeneralOrbifold}
	g = e^{2\pi iQ_L}\,\delta~,
\end{equation}
where $Q_L$ is the generator of permutations (with possible reflections) of the left-moving $T^N$ coordinates and  $\delta$ is an order $1/N$-shift in the $S^1$-direction. Of course, since we are considering an asymmetric orbifold, the construction can only take place at special points in the moduli space of the theory, where the CFT factorizes and purely left-moving lattice isometries exist. For simplicity, the models will be constructed at the fermionic point, where chiral bosons can be defined through fermionization.

We shall now define the orbifold action on the left-moving $X^I$-coordinates of $T^N$. Consider the permutations (with a possible reflection)  defined by the matrix:
\begin{equation}\label{OrbifoldAction}
	P_{IJ}(\epsilon)=\left( \begin{array}{r r r r r r}
							0 & 1 & 0 & \ldots & 0 & 0 \cr
							0 & 0 & 1 & \ldots & 0 & 0 \cr
							\vdots & \vdots & \vdots & \ldots & \vdots & \vdots \cr
							0 & 0 & 0 & \ldots & 0 & 1 \cr
							\epsilon & 0 & 0 & \ldots & 0 & 0 
						\end{array}
				\right)_{IJ}=\delta_{I,J-1}+\epsilon\,\delta_{I,J+N-1}~~,~~\textrm{where}~~\epsilon=\pm 1~,
\end{equation}
where we display the orbifold action on the relevant coordinates only. Notice that in the absence of reflection ($\epsilon=+1$), this corresponds to a $\mathbb{Z}_N$-orbifold. On the other hand, including a non-trivial reflection ($\epsilon=-1$), leads to an enhanced $\mathbb{Z}_{2N}$.

We can now pick a basis that diagonalizes $P(\epsilon)$ as follows. If $\{\lambda_I\}$ is the set of eigenvalues of $P(\epsilon)$, then we can define the linear combinations:
\begin{equation}
	Z^I =  \frac{1}{\sqrt{N}}\sum\limits_{J=1}^{N}{(\lambda_I)^{J-1}\,X^{J}}~~,~~\textrm{with}~~I=1,\ldots,N~,
\end{equation}
which are the eigenvectors of $P(\epsilon)$ with eigenvalues:
\begin{equation}
	\lambda_I(\epsilon) = e^{2\pi i (I-1+\nu)/N}~~,~~\textrm{with}~~\nu(\epsilon)=\left\{
								\begin{array}{c r}
										0 & ~,~\textrm{for}~ \epsilon=+1 \cr
										1/2 & ~,~\textrm{for}~ \epsilon=-1 
								\end{array}
	\right. ~.
\end{equation}
Notice that the new coordinates $Z^I$ can be either real or complex depending on the corresponding eigenvalue $\lambda_I$.

We now fermionize the $T^N$ coordinates of the original basis as:
\begin{equation}
	i\partial X^I = iy^I\omega^I~,
\end{equation}
where $y^I,\omega^I$ are real (auxiliary) free fermions. The change of basis is then defined by the unitary matrix:
\begin{equation}
 	U_{IJ}(\epsilon)= \frac{1}{\sqrt{N}}\left( \begin{array}{r r r r r r}
							1 & \lambda_1 & \lambda_1^2 & \ldots & \lambda_1^{N-2} & \lambda_1^{N-1} \cr
							1 & \lambda_2 & \lambda_2^2 & \ldots & \lambda_2^{N-2} & \lambda_2^{N-1} \cr
							\vdots & \vdots & \vdots & \ldots & \vdots & \vdots \cr
							1 & \lambda_N & \lambda_N^2 & \ldots & \lambda_N^{N-2} & \lambda_{N}^{N-1} 
						\end{array} \right)_{IJ} = \frac{1}{\sqrt{N}}\, e^{2\pi i(I-1+\nu)(J-1)/N}~.
\end{equation}
Note that the $\epsilon$-dependence of this matrix arises only implicitly, through the appearance of $\nu=\frac{1}{4}(1-\epsilon)$ in the eigenvalues. Explicitly, the change of basis on the coordinates and their fermionic (worldsheet) superpartners is:
\begin{eqnarray}\nonumber
	 Z^I& = &{U^I}_J(\epsilon)\,X^J~,\nonumber\\
	 \Psi^I& = &{U^I}_J(\epsilon)\,\psi^J~.
\end{eqnarray}
In this basis, the orbifold acts simply as a rotation by a phase:
\begin{eqnarray}\label{ZNbc}
	& Z^I_L \rightarrow \lambda_I(\epsilon) Z^I_L~,\nonumber \\
	& \Psi^I_L \rightarrow \lambda_I(\epsilon) \Psi^I_L~.
\end{eqnarray}
The action of the orbifold on the auxiliary free fermions is:
\begin{eqnarray}
	& y^I \rightarrow {P^I}_J(\epsilon)\,y^J~,\nonumber \\
	& \omega^I \rightarrow {P^I}_J(1)\,\omega^J~.
\end{eqnarray}
Notice that the $\omega$-fermions always transform under the orbifold as a pure permutation ($\epsilon=1$), in order for their product $y^I\omega^I$ to correctly represent the orbifold action on the bosons $\partial X^I$. Hence, the correct basis redefinition for the auxiliary free fermions is:
\begin{eqnarray}
	 Y^I &= &{U^I}_J(\epsilon)\,y^I~,\nonumber\\
	 W^I &= &{U^I}_J(1)\, \omega^J~.
\end{eqnarray}
It is, again, crucial that the $\omega$-fermions change basis as if there was no reflection ($\epsilon=+1$ or $\nu=0$).

In the new basis $\{Z^I,\Psi^I\}$, the coordinates and their fermionic superpartners generically arise as complex worldsheet fields. Complex eigenvalues always come in pairs ($\lambda^{*}=\epsilon\lambda^{N-1}$), so that we can group the associated coordinates such that $Z^{I},Z^{J}$ are complex conjugate to each other and similarly for $\Psi^{I}, \Psi^{J}$. This complexification is, in fact, crucial for the consistency of our asymmetric construction and it will permit us to represent the partition function of the theory in terms of simple level-1 characters.

It is important to note that the $Z^I$ are not fermionized into the simple product $Y^I W^I$ but, rather, into a linear combination of fermion bilinears:
\begin{eqnarray}
	 i\partial Z^I &= &i\sum\limits_{J=1}^{N}{\,{U^I}_J(\epsilon)\,{(U^{-1})^J}_K(\epsilon)\,{(U^{-1})^J}_L(1)\,Y^K\,W^L}~\nonumber\\
	& = &\frac{i}{\sqrt{N}}\,\sum\limits_{K,L=1}^{N}{\delta_{(I-K-L+1)\,{\textrm{mod}}\,N,0}\,Y^K W^L}~,
\end{eqnarray}
where the Kronecker $\delta$-function restricts the sum to those $K,L$ which satisfy the constraint $(I-K-L+1)\,{\textrm{mod}}\,N=0$. This constraint picks up precisely the combinations that correctly reproduce the $\mathbb{Z}_N$-transformation of $\partial Z^I$, as expected.

The above fermionization of the coordinates is only consistent at the fermionic point in moduli space. In our conventions, this is a square lattice $G_{IJ}=r^2\delta_{IJ}$, with $r=1/\sqrt{2}$ being the fermionic radius. The orbifold action typically introduces additional constraints. First of all, a non-trivial requirement is that the orbifold action $P(\epsilon)$ must be a symmetry of the local $\mathcal{N}=1$ superconformal theory (SCFT) on the worldsheet.
It is straightforward to check that both $T_B$ and $T_F$ of the internal SCFT are invariant, since at the fermionic point the orbifold acts crystallographically:
\begin{eqnarray}
	T^{\textrm{int}}(z)& =& -r^2\,\sum\limits_{I=1}^{N}{(\partial X^I)^2}-\frac{r^2}{2}\,\sum\limits_{I=1}^{N}{\psi^I_L\partial \psi^I_L}+\ldots = -r^2~\widehat{\sum\limits_{I,J}}{\left(\partial Z^I \partial Z^J -\frac{1}{2}\,\Psi_L^I \partial \Psi_L^J\right)}+\ldots,\nonumber\\
	T_F^{\textrm{int}}(z)& =& i\sqrt{2}\,r^2\,\sum\limits_{I=1}^{N}{\psi_L^I\, \partial X^I}+\ldots=  i\sqrt{2}\, r^2~\widehat{\sum\limits_{I,J}}{\,\Psi_L^I\,\partial Z^J}+\ldots~,
\end{eqnarray}
where  $\widehat{\sum}$ stands for the sum subject to the constraint $I+J\in N\mathbb{Z}+2(1-\nu)$ and the dots denote contributions of the remaining coordinates in $T^5$ that are not transformed, as well as the contributions due to the (super-)ghosts and the $S^1$ (super-)coordinate. Of course, at the fermionic point, $T_B$ and $T_F$ can be realized entirely in terms of the free fermions $\psi^I,y^I$ and $\omega^I$ and similar relations hold.

We can now write the modular invariant partition function of the model (up to an overall $\tau_2$-power) in the following form:
\begin{equation}\label{GeneralPartition}
	Z = \frac{1}{\eta^{12}\,\bar\eta^{12}}~R\sum\limits_{\tilde{m},n\in\mathbb{Z}}{e^{-\frac{\pi R^2}{\tau_2}|\tilde{m}+\tau n|^2}~ Z_L[^h_g](\tau)\tilde{Z}_R(\bar{\tau})\,\Gamma_{(5,5)}[^h_g](\tau,\bar\tau) }~.
\end{equation}
The orbifold twist variables are defined as follows:
\begin{equation}
	h = \frac{2\tilde p^{\mathbb{X}}}{N}~~,~~g = \frac{2 p^{\mathbb{X}}}{N}~,
\end{equation}
where $\tilde p^{\mathbb{X}}, p^{\mathbb{X}}\in\mathbb{Z}$ are the winding quantum numbers around the $S^1$, as each of the worldsheet coordinates $\sigma^1,\sigma^2$ encircle the two non-trivial cycles of the worldsheet torus.
$Z_R[^h_g]$ and $\tilde{Z}_R$ are the contributions of the left- and right- moving worldsheet fermion superpartners $\psi,\tilde{\psi}$, respectively and $\Gamma_{(5,5)}[^h_g]$ is the $(5,5)$-lattice associated to the $T^5$. The exponential factor in (\ref{GeneralPartition}) is the Lagrangian representation of the lattice partition function of the $S^1$ coordinate, $\mathbb{X}$.
It is straightforward to check that the full partition function (\ref{GeneralPartition}) is indeed modular invariant, up to an overall $\tau_2$-factor. Moreover, the partition function can be seen to arise as the modular-invariance preserving deformation of the ordinary toroidal compactification on $T^5\times S^1$. It is compatible with the constraints arising from higher-genus modular invariance and factorization and, hence, defines a class of consistent orbifold vacua.

So far, we have reviewed the construction of consistent freely-acting asymmetric $\mathbb{Z}_N$-orbifolds.
One has to keep in mind that these backgrounds are to be considered as (smooth) orbifold limits of string compactifications with  non-geometric $Q$-fluxes (T-folds).
We are now ready to examine the algebra of commutators of the internal coordinates, subject to the $\mathbb{Z}_N$-orbifold boundary conditions (\ref{ZNbc}). We will see that the asymmetric nature of the orbifold leads to a non-commutative  algebra for the  $T^N$ coordinates, the structure constants of which are nicely parametrized in terms of a flux matrix ${F^I}_J$.
We will start by writing down the mode expansions directly in the basis $Z^I$ which diagonalizes the orbifold action. We restrict our attention to the left-movers:
\begin{eqnarray}
	Z^I_L(\tau,\sigma)= \frac{i}{\sqrt{2}}\sum\limits_{k\in\mathbb{Z}}{\frac{\alpha_{k-\theta^I}^I}{k-\theta^I}~e^{-i(k-\theta^I)(\tau+\sigma)}}~,
\end{eqnarray}
where the modes satisfy the standard commutation relations:
\begin{eqnarray}\label{Acommut}
	[\alpha^I_{k-\theta^I},\alpha^J_{\ell-\theta^J}] = (k-\theta^I)\delta_{k+\ell,0}\,(UG^{-1}U^T)^{IJ}~.
\end{eqnarray}
Notice that:
\begin{eqnarray}\label{newMetric}
	(UG^{-1}U^T)^{IJ} = \frac{1}{r^2}\,\delta_{[I+J-2(1-\nu)]\,\textrm{mod}\, N,0}~,
\end{eqnarray}
 is nothing but the inverse metric in the $Z^I$-basis.
It is convenient to parametrize the orbifold action ${P^I}_J(\epsilon)$ in terms of a real antisymmetric matrix ${F^I}_J$ :
\begin{eqnarray}
		{P^I}_J={\left(e^{2\pi  F }\right)^I}_J~.
\end{eqnarray}

 The deformation angle $\theta^I=f^I \tilde p^{\mathbb{X}}$ is nothing but the $F$-flux eigenvalue $f^I$ times the winding number $\tilde p^{\mathbb{X}}$ along $S^1$ and is  related to the eigenvalue
 $\lambda_I$ through:
 \begin{eqnarray}
 	e^{2\pi i \theta^I} = \lambda_I^{\tilde p^{\mathbb{X}}}= e^{2\pi i f^I \tilde p^{\mathbb{X}}}~.
 \end{eqnarray}
Explicitly, the matrix ${F^I}_J$ can be easily constructed from its eigenvalues, $\pm i f^I$:
\begin{eqnarray}\label{Feigenvalues}
	  f^I = \frac{(I-1+\nu)}{N} \quad , \quad \textrm{with}\quad I=1, \ldots,\left[\frac{N+1}{2}\right]~,
\end{eqnarray}
and is given by the following expression:
\begin{eqnarray}\label{FQuantiz}
	{F^I}_J = -\frac{2}{N}\sum\limits_{K=1}^{\left[\frac{N+1}{2}\right]}{f^K\,\sin\left(2\pi f^K (I-J)\right)} ~,
\end{eqnarray}
where the square brackets in the upper limit of the sum denote the integer part.

 The commutation relations (\ref{Acommut}) should be supplemented with the `reality condition':
\begin{eqnarray}\label{realityCond}
	 \left.
	\begin{array}{l}
	 \overline{\alpha^I}_{k-\theta^I} = \alpha^{J}_{-k+\theta^I}   \cr
	  \theta^I= -\theta^J \\
	  \end{array}
	  \right\}  \quad ,\quad \textrm{if}~~\overline{Z^I}=Z^J~,
\end{eqnarray}
which essentially reflects the fact that only commutators between complex conjugate pairs are non-vanishing.

We can now take a pair of coordinates and evaluate their equal-times commutator using  (\ref{Acommut}), (\ref{newMetric}), and (\ref{realityCond}):
\begin{eqnarray}
	[Z^I_L(\tau,\sigma),Z^J_L(\tau,\sigma')] = \frac{1}{2}(UG^{-1}U^T)^{IJ}\, \sum\limits_{k\in\mathbb{Z}}{\frac{e^{-i(k-\theta^I)(\sigma-\sigma')}}{k-\theta^I}}~.
\end{eqnarray}
Setting $z=e^{i(\sigma'-\sigma)}$, we can express the series as the linear combination of hypergeometric functions in (\ref{hypergeometric}), which can then be analytically continued to $z\rightarrow 1$ :
\begin{eqnarray}\label{Theta}
	\Theta(\theta^I) = \Bigg\{ \begin{array}{c l}
						 -\pi \cot(\pi \theta^I) & \quad,\quad \theta^I \notin \mathbb{Z} \cr
						 0 &  \quad,\quad  \theta^I \in \mathbb{Z} \\
						\end{array}~.
\end{eqnarray}
Going back to the original basis, the commutation relations between the full coordinates $X=X_L+X_R$ become\footnote{In this asymmetric orbifold, the right-moving coordinates are
commutative. In this sense this model is different from the non-geometric flux background with elliptic monodromy, presented in section 3.1.}:
\begin{eqnarray}\label{GeneralAlgebra}
	[X^I,X^J]\Big|_{\sigma=\sigma'} = \frac{1}{2}\,\Theta(i\tilde p^{\mathbb{X}} F)^{IJ}~.
\end{eqnarray}
Note that the argument of the function $\Theta$ in (\ref{GeneralAlgebra}) is now a matrix and the easiest way to define it is through its eigenvalues. Since $F$ is at most five-dimensional and antisymmetric, it can always be expanded as:
\begin{eqnarray}\label{ThetaExp}
	\Theta(i\tilde p^{\mathbb{X}} F) = \alpha F + \beta F^3~,
\end{eqnarray}
with $\alpha,\beta$ complex coefficients that are determined by substituting in the eigenvalues of $F$, eq. (\ref{Feigenvalues}).

In the cases $\mathbb{Z}_2,\mathbb{Z}_3$ and $\mathbb{Z}_4$, there is only a single complex eigenvalue\footnote{In the $\mathbb{Z}_4$ case with $N=2$, $\epsilon=-1$, the complex eigenvalue comes with multiplicity two because of the doubling of the orbifold action on $T^4$.} and, hence, $\beta=0$. For the $\mathbb{Z}_k$ orbifold with $k=2,3,4$, the non-commutative algebra between the toroidal coordinates then takes the simple form:
\begin{eqnarray}\label{Z234commut}
	 [X^I,X^J] = \frac{i}{2} k\, F^{IJ}\,\Theta\left( \tilde p^{\mathbb{X}}/k\right) = \left\{ \begin{array}{c l}
	 						 -\frac{i}{2} \pi k\, F^{IJ}\,\cot\left(\frac{\pi \tilde p^{\mathbb{X}}}{k}\right)  & ,\quad \tilde p^{\mathbb{X}} \notin k\mathbb{Z} \cr
							 0 & ,\quad \tilde p^{\mathbb{X}} \in  k\mathbb{Z} \\
	 				\end{array}
	 \right.	 ~.
\end{eqnarray}
The case of $\mathbb{Z}_2$ is special, because the r.h.s. vanishes identically and the toroidal coordinates remain commutative. This is, essentially, due to the fact that the $\mathbb{Z}_2$-action can always be reduced to a reflection of a real coordinate and, hence, it does not really `entangle' different coordinates.

It is instructive to comment on the structure of the general result (\ref{GeneralAlgebra}). First of all, the stringy origin of the non-commutative behavior becomes manifest, by noticing that $[X^I,X^J]$  is always proportional to the string scale $\alpha'$. Secondly, the explicit periodic dependence on the winding number $\tilde p^{\mathbb{X}}$ along the circle direction $\mathbb{X}$, implies that only when non-trivial winding modes along $S^1$ can be excited, can one probe the non-commutativity in the toroidal coordinates. 

\section{Conclusions}

As we have seen, non-geometric closed string backgrounds lead to spatial non-commutativity and non-associativity of the closed string coordinates. The associated algebras 
can be described by twisted Poisson structures in analogy to the momentum phase space of point particles in the field of magnetic monopoles. In string theory, the deformation parameters
of the non-commutative/non-associative algebras are given in terms of tensors ($R$-flux) and connections ($Q$-flux), which naturally arise  in the context of doubled field theory and effective
actions of non-geometric backgrounds. Since closed strings are responsible for gravity, it is tempting to conjecture that the non-commmutative/non-associative closed string geometries
will lead to a non-commmutative/non-associative deformation of gravity. Furthermore, as these non-associative algebras go beyond  matrix (quantum) mechanics, it is an interesting question, 
how to represent them, possibly by some kind of octonionic objects.

\section*{Acknowledgments}
\vspace{-0.25cm}
We would like to thank D. Andriot, R. Blumenhagen, C. Condeescu, A. Deser, I. Florakis, O. Hohm, M. Larfors, P. Patalong E. Plauschinn and F. Renneke for the very pleasant collaborations
on the material presented in this paper.
For very useful comments and discussions we would like to thank I. Bakas and T. Strobl.
This work is supported by  the
DFG Transregional Collaborative Research Centre TRR 33
and the DFG cluster of excellence `Origin and Structure of the Universe'.


\begin{thebibliography}{99}





\bibitem{Grana:2005jc}
  M.~Grana,
  ``Flux compactifications in string theory: A Comprehensive review,''
  Phys.\ Rept.\  {\bf 423} (2006) 91
  [hep-th/0509003].


\bibitem{Blumenhagen:2006ci}
  R.~Blumenhagen, B.~K\"ors, D.~L\"ust and S.~Stieberger,
  ``Four-dimensional String Compactifications with D-Branes, Orientifolds and Fluxes,''
  Phys.\ Rept.\  {\bf 445} (2007) 1
  [hep-th/0610327].
  
  \bibitem{Chu:1998qz}
C.-S. Chu and P.-M. Ho, {\it {Noncommutative open string and D-brane}},  {\em
  Nucl.Phys.} {\bf B550} (1999) 151--168,
  [\href{http://xxx.lanl.gov/abs/hep-th/9812219}{{\tt hep-th/9812219}}].

\bibitem{Schomerus:1999ug}
V.~Schomerus, {\it {D-branes and deformation quantization}},  {\em JHEP} {\bf
  9906} (1999) 030, [\href{http://xxx.lanl.gov/abs/hep-th/9903205}{{\tt
  hep-th/9903205}}].

\bibitem{Seiberg:1999vs}
N.~Seiberg and E.~Witten, {\it {String theory and noncommutative geometry}},
  {\em JHEP} {\bf 9909} (1999) 032,
  [\href{http://xxx.lanl.gov/abs/hep-th/9908142}{{\tt hep-th/9908142}}].

\bibitem{Ardalan:1998ks}
F.~Ardalan, H.~Arfaei, and M.~Sheikh-Jabbari, {\it {Mixed branes and M(atrix)
  theory on noncommutative torus}},
  \href{http://xxx.lanl.gov/abs/hep-th/9803067}{{\tt hep-th/9803067}}.

\bibitem{Hull:2004in}
  C.~M.~Hull,
  ``A Geometry for non-geometric string backgrounds,''
  JHEP {\bf 0510}, 065 (2005)
  [hep-th/0406102].

\bibitem{Shelton:2005cf}
  J.~Shelton, W.~Taylor and B.~Wecht,
  ``Nongeometric flux compactifications,''
  JHEP {\bf 0510} (2005) 085
  [hep-th/0508133].

\bibitem{Dabholkar:2005ve}
  A.~Dabholkar and C.~Hull,
  ``Generalised T-duality and non-geometric backgrounds,''
  JHEP {\bf 0605}, 009 (2006)
  [hep-th/0512005].


\bibitem{Blumenhagen:2010hj}
  R.~Blumenhagen, E.~Plauschinn,
  ``Nonassociative Gravity in String Theory?,''
  J.\ Phys.\ A {\bf A44 } (2011)  015401.
  [arXiv:1010.1263 [hep-th]].

\bibitem{Lust:2010iy}
  D.~L\"ust,
  ``T-duality and closed string non-commutative (doubled) geometry,''
  JHEP {\bf 1012 } (2010)  084.
  [arXiv:1010.1361 [hep-th]].

  
\bibitem{Blumenhagen:2011ph}
  R.~Blumenhagen, A.~Deser, D.~L\"ust, E.~Plauschinn and F.~Rennecke,
  ``Non-geometric Fluxes, Asymmetric Strings and Nonassociative Geometry,''
  J.\ Phys.\ A A {\bf 44} (2011) 385401
  [arXiv:1106.0316 [hep-th]].
  
  
  
\bibitem{Blumenhagen:2011yv}
  R.~Blumenhagen,
  ``Nonassociativity in String Theory,''
  arXiv:1112.4611 [hep-th].
  
  
\bibitem{Condeescu:2012sp}
  C.~Condeescu, I.~Florakis and D.~L\"ust,
  ``Asymmetric Orbifolds, Non-Geometric Fluxes and Non-Commutativity in Closed String Theory,''
  arXiv:1202.6366 [hep-th].

  
   
\bibitem{Plauschinn:2012kd}
  E.~Plauschinn,
  ``Non-geometric fluxes and non-associative geometry,''
  arXiv:1203.6203 [hep-th].
  
\bibitem{Saemann:2012ex}
  C.~Saemann and R.~J.~Szabo,
  ``Groupoid Quantization of Loop Spaces,''
  arXiv:1203.5921 [hep-th].

\bibitem{Jackiw:1984rd}
R.~Jackiw, {\it {3 - Cocycle in Mathematics and Physics}},  {\em
  Phys.Rev.Lett.} {\bf 54} (1985) 159--162.


\bibitem{Klimcik:2001vg}
  C.~Klimcik, T.~Strobl,
  ``WZW - Poisson manifolds,''
  J.\ Geom.\ Phys.\  {\bf 43 } (2002)  341-344.
  [math/0104189 [math-sg]].
  
\bibitem{Alekseev:2004np}
  A.~Alekseev, T.~Strobl,
  ``Current algebras and differential geometry,''
  JHEP {\bf 0503 } (2005)  035.
  [hep-th/0410183].

\bibitem{Kotov:2004wz}
  A.~Kotov, P.~Schaller, T.~Strobl,
  ``Dirac sigma models,''
  Commun.\ Math.\ Phys.\  {\bf 260 } (2005)  455-480.
  [hep-th/0411112].
  
  
  
  
\bibitem{Hull:2006va}
  C.M. Hull,
  ``Doubled Geometry and T-Folds,''
  JHEP {\bf 0707} (2007) 080
  [hep-th/0605149].
  
\bibitem{Hull:2009mi}
  C.~Hull and B.~Zwiebach,
  ``Double Field Theory,''
  JHEP {\bf 0909} (2009) 099
  [arXiv:0904.4664 [hep-th]].


\bibitem{Hull:2009zb}
  C.~Hull and B.~Zwiebach,
  ``The Gauge algebra of double field theory and Courant brackets,''
  JHEP {\bf 0909} (2009) 090
  [arXiv:0908.1792 [hep-th]].


\bibitem{Hohm:2010jy}
  O.~Hohm, C.~Hull and B.~Zwiebach,
  ``Background independent action for double field theory,''
  JHEP {\bf 1007} (2010) 016
  [arXiv:1003.5027 [hep-th]].

  
\bibitem{Hohm:2010pp}
  O.~Hohm, C.~Hull and B.~Zwiebach,
  JHEP {\bf 1008} (2010) 008
  [arXiv:1006.4823 [hep-th]].
 


\bibitem{Andriot:2011uh}
  D.~Andriot, M.~Larfors, D.~L\"ust and P.~Patalong,
  ``A ten-dimensional action for non-geometric fluxes,''
  JHEP {\bf 1109} (2011) 134
  [arXiv:1106.4015 [hep-th]].


\bibitem{Andriot:2012wx}
  D.~Andriot, O.~Hohm, M.~Larfors, D.~L\"ust and P.~Patalong,
  ``A geometric action for non-geometric fluxes,'' to appear in Phys.Rev.Lett.,
  arXiv:1202.3060 [hep-th].


\bibitem{Patalong:2012np}
  P.~Patalong,
  ``Non-geometric Q-flux in ten dimensions,''
  arXiv:1203.5127 [hep-th].
  
\bibitem{Larfors:2012zz}
  M.~Larfors,
  ``Non-geometric fluxes in ten dimensions,''


\bibitem{Andriot:2012an}
  D.~Andriot, O.~Hohm, M.~Larfors, D.~L\"ust and P.~Patalong,
  ``Non-Geometric Fluxes in Supergravity and Double Field Theory,''
  arXiv:1204.1979 [hep-th].
  
  
  
  
  
\bibitem{Blumenhagen:2012ma}
  R.~Blumenhagen, A.~Deser, E.~Plauschinn and F.~Rennecke,
  ``Palatini-Lovelock-Cartan Gravity - Bianchi Identities for Stringy Fluxes,''
  arXiv:1202.4934 [hep-th].
  
  
  \bibitem{Bouwknegt:2000qt}
P.~Bouwknegt and V.~Mathai, {\it {D-branes, B fields and twisted K theory}},
  {\em JHEP} {\bf 0003} (2000) 007,
  [\href{http://xxx.lanl.gov/abs/hep-th/0002023}{{\tt hep-th/0002023}}].
  
  \bibitem{Bouwknegt:2004ap}
P.~Bouwknegt, K.~Hannabuss, and V.~Mathai, {\it {Nonassociative tori and
  applications to T-duality}},  {\em Commun.Math.Phys.} {\bf 264} (2006)
  41--69, [\href{http://xxx.lanl.gov/abs/hep-th/0412092}{{\tt
  hep-th/0412092}}].

\bibitem{Mathai:2004qq}
V.~Mathai and J.~M. Rosenberg, {\it {T duality for torus bundles with H fluxes
  via noncommutative topology}},  {\em Commun.Math.Phys.} {\bf 253} (2004)
  705--721, [\href{http://xxx.lanl.gov/abs/hep-th/0401168}{{\tt
  hep-th/0401168}}].

\bibitem{Mathai:2004qc}
V.~Mathai and J.~M. Rosenberg, {\it {On Mysteriously missing T-duals, H-flux
  and the T-duality group}},
  \href{http://xxx.lanl.gov/abs/hep-th/0409073}{{\tt hep-th/0409073}}.
  
  
\bibitem{Grange:2006es}
  P.~Grange and S.~Schafer-Nameki,
  ``T-duality with H-flux: Non-commutativity, T-folds and G x G structure,''
  Nucl.\ Phys.\ B {\bf 770} (2007) 123
  [hep-th/0609084].
  
  
\bibitem{Grange:2007bp}
  P.~Grange and S.~Schafer-Nameki,
  ``Towards mirror symmetry a la SYZ for generalized Calabi-Yau manifolds,''
  JHEP {\bf 0710} (2007) 052
  [arXiv:0708.2392 [hep-th]].
  
  


\bibitem{Brodzki:2007hg}
J.~Brodzki, V.~Mathai, J.~M. Rosenberg, and R.~J. Szabo, {\it {Noncommutative
  correspondences, duality and D-branes in bivariant K-theory}},  {\em
  Adv.Theor.Math.Phys.} {\bf 13} (2009) 497--552,
  [\href{http://xxx.lanl.gov/abs/0708.2648}{{\tt arXiv:0708.2648}}].




\end{thebibliography}
\end{document}